\documentstyle[prl,aps,floats]{revtex}
\input{epsf}

\newcommand{\be}{\begin{equation}}
\newcommand{\ee}{\end{equation}}
\newcommand{\bea}{\begin{eqnarray}}
\newcommand{\eea}{\end{eqnarray}}
\newcommand{\bean}{\begin{eqnarray*}}

\newcommand{\eean}{\end{eqnarray*}}
\font\upright=cmu10 scaled\magstep1
\font\sans=cmss10
\newcommand{\ssf}{\sans}
\newcommand{\stroke}{\vrule height8pt width0.4pt depth-0.1pt}
\newcommand{\Z}{\hbox{\upright\rlap{\ssf Z}\kern 2.7pt {\ssf Z}}}

\newcommand{\C}{{\rlap{\rlap{C}\kern 3.8pt\stroke}\phantom{C}}}
\newcommand{\R}{\hbox{\upright\rlap{I}\kern 1.7pt R}}
\newcommand{\CP}{\C{\upright\rlap{I}\kern 1.5pt P}}
\newcommand{\PP}{\hbox{\upright\rlap{I}\kern 1.5pt P}}

\newcommand{\identity}{{\upright\rlap{1}\kern 2.0pt 1}}

\newcommand{\HH}{\mbox{\hbox{\upright\rlap{I}\kern 1.7pt H}}}

\font\mybb=msbm10 at 11pt

\def\bb#1{\hbox{\mybb#1}}

\def\bC {\bb{C}}

\renewcommand{\CP}{\bC {\rm P}}

\tighten
\begin{document} 
\draft
\twocolumn[\hsize\textwidth\columnwidth\hsize\csname@twocolumnfalse\endcsname
\preprint{Imperial/TP/96-97/51}
\title{Solitonic fullerene structures in light atomic nuclei}
\author{Richard A. Battye{$^{1}$} and Paul M. Sutcliffe{$^{2}$}}
\address{${}^1$ Department of Applied Mathematics and Theoretical
Physics, Centre for Mathematical Sciences, University of Cambridge, \\
Wilberforce Road, Cambridge CB3 OWA, U.K. \\ 
${}^2$ Institute of Mathematics, University of Kent at Canterbury,
Canterbury CT2 7NF, U.K.
}
\maketitle
\begin{abstract}
The Skyrme model is a classical field theory which has topological
soliton solutions. These solitons are candidates for describing
nuclei, with an identification between the numbers of solitons and
nucleons.  We have computed numerically, using two different
minimization algorithms, minimum energy configurations for up to 22
solitons. We find, remarkably, that the solutions for seven or more
solitons have nucleon density isosurfaces in the form of polyhedra
made of hexagons and pentagons.  Precisely these structures arise,
though at the much larger molecular scale, in the chemistry of carbon
shells, where they are known as fullerenes.
\end{abstract}

\date{\today}

\pacs{PACS Numbers : }
]

\renewcommand{\thefootnote}{\arabic{footnote}}
\setcounter{footnote}{0}

The Skyrme model~\cite{Sk} was first proposed in the early sixties as
 a model for the strong interactions of hadrons, but it was set aside
 after the advent of quantum chromodynamics (QCD). Much later
 Witten~\cite{WIT} showed that it could arise as an effective 
 description at low energies in the limit where the number of quark
 colours is large. Subsequent work~\cite{ANW} demonstrated that the
 single soliton solution (known as a Skyrmion)  reproduced the
 properties of a nucleon to within an accuracy of around $30\%$;
 quite an achievement, given that there is, at present, no
 practical way of calculating the properties of nuclei from QCD via,
 for example, lattice gauge theory. 

In order to study nuclei of larger atomic number one first needs to
compute the minimal energy configurations of multi-solitons, since in
the Skyrme model there is an identification between the numbers of
solitons and nucleons.  Here, we present the results of an extensive
set of simulations using two very different approaches designed to
compute the minimal energy solutions for upto 22 solitons. With a
small number of caveats, these results establish an attractive
analogy with fullerene cages familiar in carbon
chemistry~\cite{kroto,atlas}. Although these classical solutions must first be quantized before a
final comparison with experimental data can be performed it is
expected that  quantum corrections will be relatively small, since we
are dealing with solitons, and so the classical solutions will contain
important physically relevant information about the properties of
nuclei. 

The Skyrme model is defined in terms of an $SU(2)$ valued field $U({\bf x})$, 
with an associated static energy  
\bea
E =  \frac{1}{24\pi^2}
\int
\bigg\{ & &{\rm Tr}(
\partial_iU\partial_iU^{-1})\cr
& &-{1\over 8} {\rm Tr}\left([(\partial_iU) U^{-1},(\partial_jU)
U^{-1}]^2\right)\bigg\}\ d^3{\bf x}\,.
\label{energy}
\eea
Note that the two physically relevant constants which would appear in front of each
of the two terms in the most general version of
(\ref{energy}) have, for convenience, been scaled out by an 
appropriate choice of energy and length units.
For finite energy, we impose the boundary condition $U(\infty)=1,$ and
pions are described by the usual quantum field theory treatment
of fluctuations of the Skyrme field around this vacuum value, 
but nucleons arise in a very different manner, as classical soliton
solutions.

The boundary condition implies a compactification of
the domain, and therefore $U$ is a map from 
compactified $\R^3\sim S^3\mapsto S^3,$ since $S^3$ is the
manifold of the target space, the group $SU(2).$
 Such mappings have  non-trivial homotopy classes characterized by an
integer valued winding number, which
has the explicit representation
\be
B=-{\epsilon_{ijk}\over 24\pi^2}\int\
 {\rm Tr}\left((\partial_iU) U^{-1} (\partial_jU) U^{-1} (\partial_kU) U^{-1}\right)\ d^3{\bf x}\,.
\label{baryon}
\ee  
$B$, which stands for baryon number or the number of nucleons,
is often referred to as the topological charge and is
the number of solitons in a given field configuration.
A simple manipulation of equations (\ref{energy}) and (\ref{baryon})
allows one to deduce the Faddeev-Bogomolny (FB) bound $E\ge|B|$.
Generically, a charge
$B$ field will have an energy density ${\cal E}$ 
(the integrand of (\ref{energy})) and a baryon density ${\cal B}$ (the
integrand of (\ref{baryon})) both of which consists of $B$ well-separated
lumps localized in space. However, as we discuss below, this is not
the case for the minimal energy fields in which the solitons are close
together.

The mathematical problem is to find, for each integer $B$, the
field $U$ which minimizes the energy (\ref{energy}) subject to the
constraint (\ref{baryon}).  This can be addressed by numerical
algorithms designed to minimize either a  discretized version of the
energy (\ref{energy}) or, equivalently, by solving a  discretized
version of the second-order field equations which follow from the
variation of (\ref{energy}). This first approach 
is a very demanding computational
exercise (see ref.~\cite{BS5} for a detailed discussion),
 requiring the use of modern parallel supercomputers, but
results for low charge ($B\le 8$) were found using this 
method~\cite{KS,V,BTC,BS2}. The results presented here
extend this numerical approach up to $B=22.$ In addition, we have
applied a second, very different, technique to the construction of
minimal energy solitons which not only
allows us to have greater confidence that the numerical solutions we
have constructed are indeed the global minima, but in addition
provides a good analytic approximation to the numerical solutions,
making it much easier to identify their structure and symmetries.

\begin{figure}
\begin{center}
\epsfxsize=9cm\epsfysize=20cm\epsffile{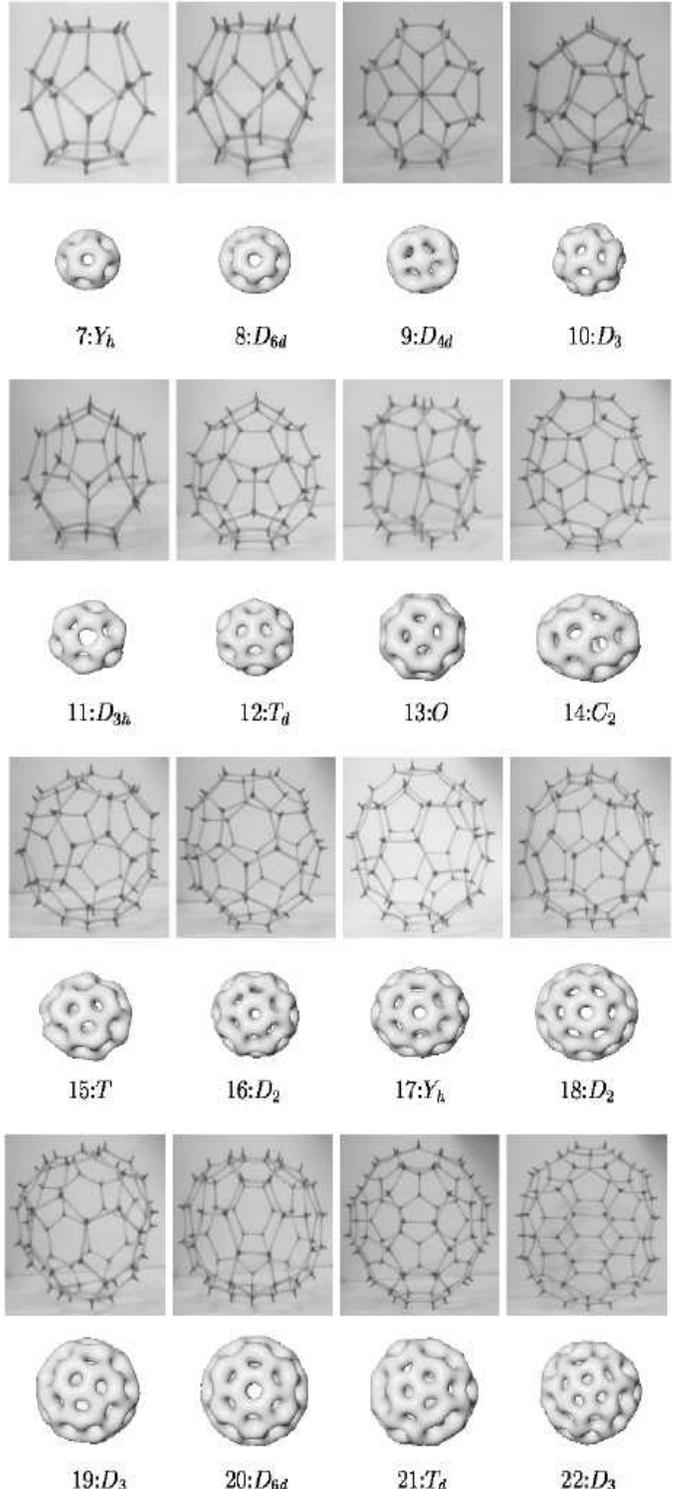}
\caption{The baryon density isosurfaces for
the solutions which we have identified as the minima for $7\le B\le
22$, and the associated polyhedral models.
 The isosurfaces correspond to ${\cal B}=0.035$ and are presented to
scale, whereas the polyhedra are not to scale.}    
\label{fig-solutions}
\end{center}
\end{figure}

Our second approach makes use of the remarkable fact that minimal
energy Skyrmions can be approximated by an ansatz involving rational
maps between Riemann spheres~\cite{HMS}; a result which we will
further confirm.
The Skyrme field is a map $U:\R^3\mapsto
S^3$, so it is not immediately obvious how to obtain such a map from 
rational maps which are 
between spheres $S^2\mapsto S^2.$ Briefly, the domain $S^2$ of the
rational map is identified with concentric spheres in $\R^3$, and the
target $S^2$ with spheres of latitude on $S^3.$  To present the ansatz
it is convenient to use spherical coordinates in $\R^3$, so that a
point ${\bf x}\in\R^3$ is given  by a pair $(r,z)$, where $r=\vert{\bf
x}\vert$ is the  distance from the origin, and $z$ is a Riemann sphere
coordinate, namely $z=\tan[\theta/2]\exp[i\phi]$ where $\theta$ and
$\phi$ are the normal spherical polar coordinates.

Now, let $R(z)$ be a degree $B$ rational map,
that is, $R=p/q$ where $p$ and $q$ are polynomials in $z$ such that
$\max[\mbox{deg}(p),\mbox{deg}(q)]=B$, with  no common
factors.  Given such a rational map the ansatz for the Skyrme field is
\be  
U(r,z)=\exp\bigg[\frac{if(r)}{1+\vert R\vert^2} \pmatrix{1-\vert
R\vert^2& 2\bar R\cr 2R & \vert R\vert^2-1\cr}\bigg]\,,
\label{rma}
\ee
where $f(r)$ is a real profile function satisfying the  boundary
conditions $f(0)=\pi$ and $f(\infty)=0$. This is determined by
minimization of the Skyrme energy of the field (\ref{rma}) given a
particular rational map $R$. The ansatz yields an exact solution
for $B=1$ and it was shown in
ref.\cite{HMS} that for $2\le B\le 8$, suitable maps exist for which
the field (\ref{rma}) is a good approximation to the numerically
computed solutions, in the sense that the symmetry is identical
and the energy
is only one or two percent above the numerically computed values.

Substituting the ansatz (\ref{rma}) into the energy (\ref{energy})
produces an energy function on the space of rational maps, which we
denote by ${\cal I}(R)$, given by 
\be 
{\cal I}(R)=\frac{1}{4\pi}\int \bigg(
\frac{1+\vert z\vert^2}{1+\vert R\vert^2}
\bigg\vert\frac{dR}{dz}\bigg\vert\bigg)^4 \frac{2i \  dz  d\bar z
}{(1+\vert z\vert^2)^2}\,.
\label{i}
\ee
Therefore, our second approach to computing minimal
energy Skyrmions is to search
the (finite dimensional) parameter space of general degree $B$
rational maps to find the one which minimizes
${\cal I}(R),$ using a powerful numerical minimization 
technique known as simulated annealing~\cite{sabook}. 

Clearly, this procedure is not guaranteed to find the minimum energy
Skyrmion since the topography of the rational map space may be slightly
different to that of the full non-linear field theory, but as we shall
see for the most part it works well, only encountering difficulties
 when there are two
or more Skyrmion solutions, either saddle points or genuine local
minima, with very similar energies.  We use the first minimization
technique as a check, and in the small number of  cases where relaxing
well-separated clusters consistently yields a different solution
for a wide range of initial conditions, the symmetry of the Skyrmion
solution is identified by eye from the baryon density isosurface, and
an approximate rational map can then be found by relaxing in the
rational map space restricted to have the correct symmetry.  In such
cases the values of ${\cal I}$ for the different solutions are usually
very close~\cite{BS5}.

The results of applying the two minimization techniques in this way
are presented  in table~\ref{tab-solutions} for $1\le B\le 22$
and pictorially in Fig.~\ref{fig-solutions} for $7\le B\le 22$. In all
but a small number of cases ($B=10,14,16,22$) we find that the minimum
energy Skyrmion and that in the rational map space
coincide. Furthermore, for each of these special cases, except
$B=14$, we were able to find a map with the same symmetry.
 For $B=14$ we were prevented from finding a rational
map approximation for the true minimum  since its symmetry group, $C_2$, is
a subgroup of that of the minimum energy rational map, $D_2$.

The baryon density isosurface can be associated with a polyhedron
whose edges and vertices coincide with the 
regions in which the baryon density is localized.
  Examination of the solutions shows that, with the  exception
of $B=9$ and $13$, the associated polyhedra are trivalent with
$4(B-2)$ vertices (the Geometric Energy Minimization (GEM)  rules) as
predicted in ref.~\cite{BS2}, and for $B\ge 7$ they  comprise
of 12 pentagons and $2(B-7)$ hexagons. Such structures are common in a
wide range of physical applications, and have become a hot
topic in carbon chemistry where they correspond to shells with
carbon atoms placed at the vertices, the most famous being the
icosahedrally symmetric Buckminsterfullerene $C_{60}$ structure, which
is also the traditional soccer ball design.  For this reason
we shall refer to such solutions as being of the fullerene type, with
the prediction, spectacularly confirmed by our results in all cases
except $B=9$ and $B=13$, that the polyhedron associated with the
Skyrmion of charge $B$, has a structure from the family
of carbon cages for $C_{4(B-2)}$. 

We had predicted in ref.~\cite{BS2} that the Buckyball $C_{60}$ configuration
would be found for $B=17$ and indeed an approximate rational map description
was found in ref.~\cite{HMS}. Here, we see that such a solution is the
minimum energy solution of the full non-linear field equations and in
the rational map space. We see also that a large number of the other
solutions  have platonic symmetries which, from the mechanical point
of view, implies the structure packs well. It would appear, therefore, that
such structures may be preferred over less symmetric ones in the
minimization procedure. We should note, however, that this is not
always the case and  rational maps with platonic symmetries can easily
be found, for example at $B=9$, which do not give minima~\cite{BS5}.

The polyhedra found for $B=9$ and $13$ do not obey the GEM rules, nor
are they of the fullerene type, since they contain four-valent
links. They can, however, be related to a fullerene via the concept of
symmetry enhancement, as follows. A very common structure within the fullerene
polyhedra is two pentagons separated by two hexagons. If the edge
which is common to the two hexagons is shrunk to have zero length, the
four polygons then form a $C_4$ symmetric configuration containing a
four valent bond. For the case of $B=9$, the polyhedron can be thought
of as being created from a $D_2$ symmetric fullerene by the action  of
two such operations, and in the $B=13$ case six operations can be used to convert
another $D_2$ configuration into  one with  $O$ symmetry. Empirically,
we see that each symmetry enhancement operation appears to be
accompanied by an equivalent one antipodally placed on the
polyhedron, and single operations appear not to occur.

\begin{table}
\centering
\begin{tabular}{|c|c|c|c|c|c|}
\hline $B$ & $G$ & $E/B$ & $E_B$ & $I_B$ & $\Delta E/B$ \\ \hline  
    1 & $O(3)$ &   1.2322 &   1.2322 &   0.0000 &   0.0000 \\
    2 & $D_{\infty h}$ &   1.1791 &   2.3582 &   0.1062 &   0.0531 \\
    3 & $T_d$ &   1.1462 &   3.4386 &   0.1518 &   0.0860 \\
    4 & $O_h$ &   1.1201 &   4.4804 &   0.1904 &   0.1121 \\
    5 & $D_{2d}$ &   1.1172 &   5.5860 &   0.1266 &   0.1150 \\
    6 & $D_{4d}$ &   1.1079 &   6.6474 &   0.1708 &   0.1243 \\
    7 & $Y_h$ &   1.0947 &   7.6629 &   0.2167 &   0.1375 \\
    8 & $D_{6d}$ &   1.0960 &   8.7680 &   0.1271 &   0.1362 \\
    9 & $D_{4d}$ &   1.0936 &   9.8424 &   0.1578 &   0.1386 \\
   10 & $D_3$ &   1.0904 &  10.9040 &   0.1706 &   0.1418 \\
   11 & $D_{3h}$ &   1.0889 &  11.9779 &   0.1583 &   0.1433 \\
   12 & $T_d$ &   1.0856 &  13.0272 &   0.1829 &   0.1466 \\
   13 & $O$ &   1.0834 &  14.0842 &   0.1752 &   0.1488 \\
   14 & $C_2$ &   1.0842 &  15.1788 &   0.1376 &   0.1480 \\
   15 & $T$ &   1.0825 &  16.2375 &   0.1735 &   0.1497 \\
   16 & $D_2$ &   1.0809 &  17.2944 &   0.1753 &   0.1513 \\
   17 & $Y_h$ &   1.0774 &  18.3158 &   0.2108 &   0.1548 \\
   18 & $D_2$ &   1.0788 &  19.4184 &   0.1296 &   0.1534 \\
   19 & $D_3$ &   1.0786 &  20.4934 &   0.1572 &   0.1536 \\
   20 & $D_{6d}$ &   1.0779 &  21.5580 &   0.1676 &   0.1543 \\
   21 & $T_d$ &   1.0780 &  22.6380 &   0.1522 &   0.1542 \\
   22 & $D_3$ &   1.0766 &  23.6852 &   0.1850 &   0.1556 \\
\hline
\end{tabular}
\caption{A summary of the symmetries and energies of the Skyrmion
configurations which we have identified as the minima.  Included
is the ionization energy ($I_B$) --- that required to remove one
Skyrmion --- and the binding energy per Skyrmion ($\Delta E/B$)  ---
that  required to split the charge $B$ Skyrmion into $B$ charge
one Skyrmions divided by the total number.}
\label{tab-solutions}
\end{table}

We have computed the energies of the solutions which are presented in
table~\ref{tab-solutions} using the rational map ansatz to create
initial conditions which are then relaxed under the action of the full
non-linear field equations. It should be noted that these values are
(for $B>1$) always a little less than the corresponding values computed
solely within the rational map ansatz.  On the discrete grid the
computed value of the baryon number, $B_{\rm dis}$, is less than the
corresponding integer $B$ suggesting that the finite difference
approximations used to compute the energy $E_{\rm dis}$
will under-estimate the true energy. Moreover, in
the initial conditions one must impose the boundary condition $U=1$ at
the edge of the box. By using a wide range of different grid sizes and
spacing we have shown~\cite{BS5} that the value of $E_{\rm dis}/B_{\rm
dis}$ can be computed accurately, and hence so can the true energy
using the formula
$E_B=B*(E_{\rm dis}/B_{\rm dis})$. We claim that our determinations of
$E_{\rm dis}/B_{\rm dis}$ are accurate in the absolute sense to within
$\pm 0.001$, and that the relative values are probably even more
accurate.

\begin{figure}
\centerline{\epsfxsize=5cm\epsfysize=4.3cm\epsffile{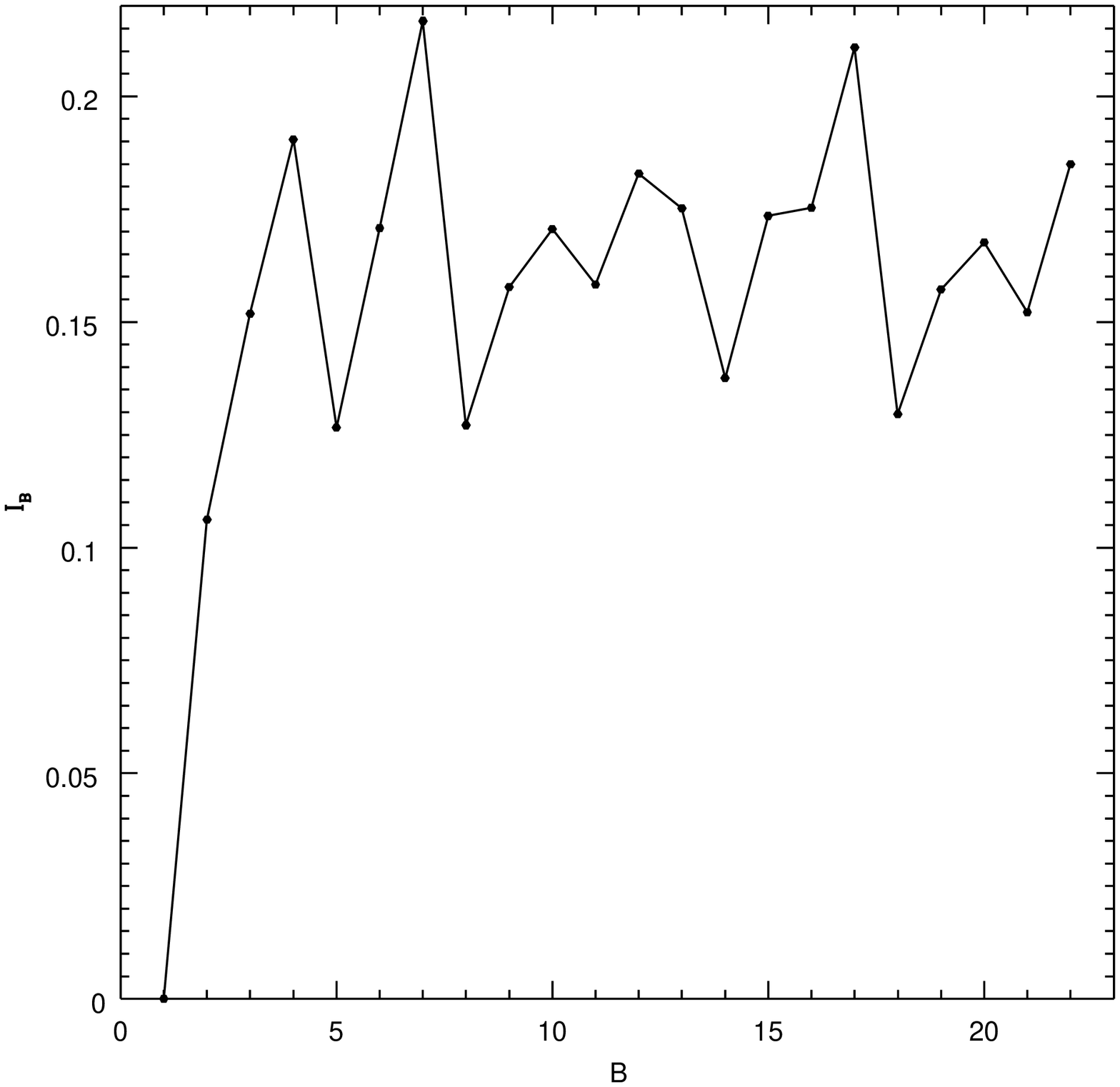}}
\centerline{\epsfxsize=5cm\epsfysize=4.3cm\epsffile{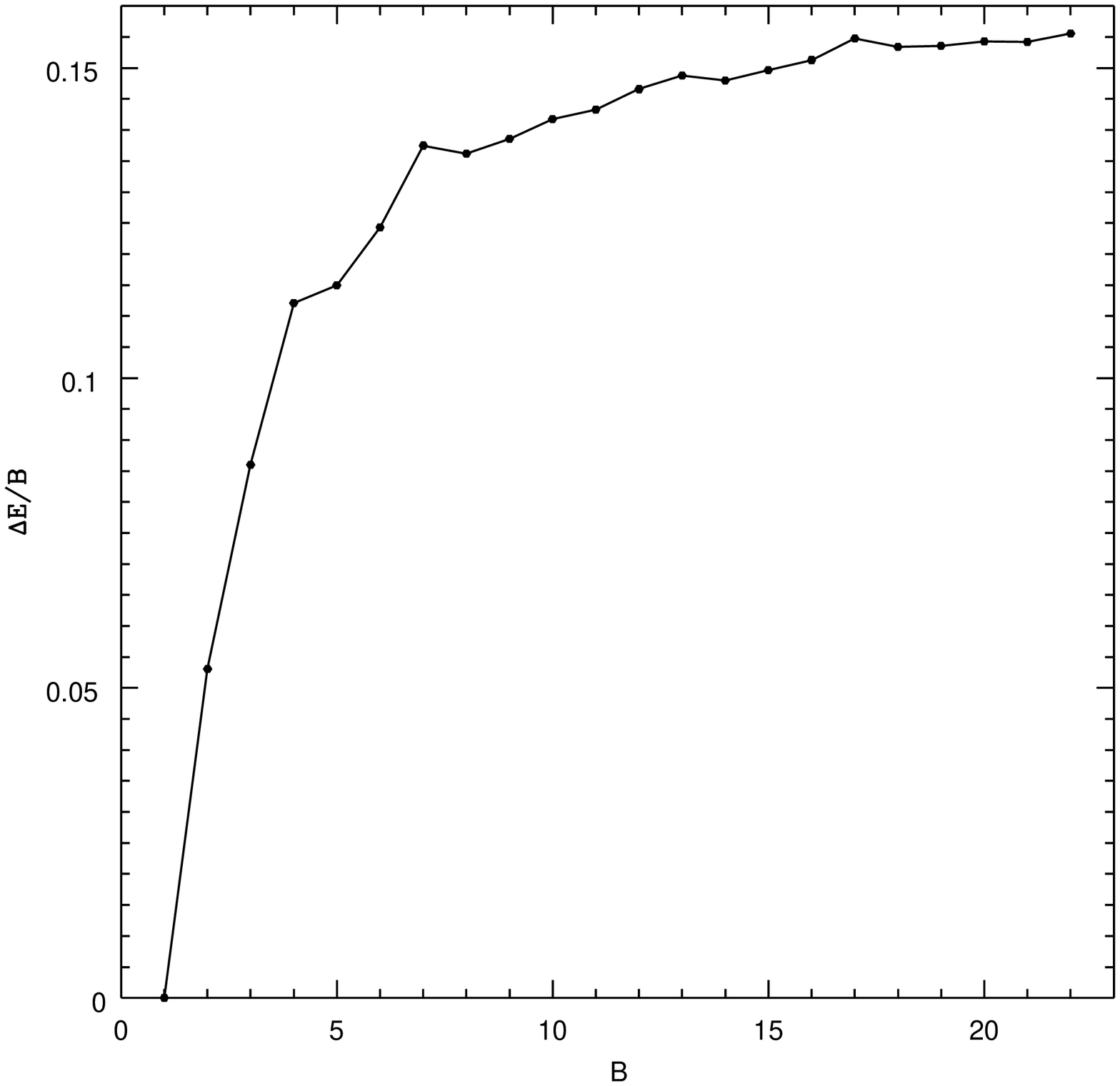}}
\caption{On top the ionization energy $I_B$ plotted against $B$.
Notice that the most stable
solutions are those with the most symmetry, $B=4,7,17$, while the
least stable are those with little symmetry $B=5,8,14,18$. On the 
bottom the binding energy per baryon $\Delta E/B$ plotted
against $B$. We see that for large $B$ the binding energy appears
to level out at around $0.15-0.16$ as one might expect in a simple
model of nuclei.}    
\label{fig-ibind}
\end{figure}

We have also computed the ionization energy $I_B=E_{B-1}+E_1-E_B$,
which is the energy required to remove a single Skyrmion, and the
binding energy per nucleon $\Delta E/B=E_1-(E/B)$ which is the energy
required to separate the solution into $B$ well-separated Skyrmions
divided by the total baryon number. These values are tabulated in
table~\ref{tab-solutions} and are plotted against $B$ 
in Fig.~\ref{fig-ibind}. The ionization energy is largest for the most
symmetrical solutions $B=4,7$ and $17$, and is least
for those with little symmetry,
$B=5,8,14$ and $18$, which is very much as one would expect. The
binding energy appears to increase to an asymptotic value of around $0.15 -
0.16$. This is a clear consequence of the FB bound since it is
linearly related to $E/B$.

In fact the value of $E/B$ appears to have an asymptotic value which is
around $6\%-7\%$ above the FB bound, compatible with the value
obtained for a hexagonal lattice~\cite{BS3} which is the limit of an
infinitely large fullerene (the analogue of graphite in carbon
chemistry). It is clear that an infinitely large shell is physically
unlikely and that there probably exists a value $B_*$ such that for
$B>B_*$ the solutions no longer look like fullerene shells. In such a
case the solutions are likely to begin to look more like portions cut
from the infinite Skyrme crystal~\cite{crystal} whose $E/B$ is only
$4\%$ above the FB bound. Another possibility is an intermediate state
comprising of multiple shells~\cite{MP}, although all the known
configurations of this kind have much larger values of $E/B$.
We have definitely shown that $B_*> 22$ and we believe that the connection
between Skyrmions, fullerenes and rational maps will continue for much
larger values of $B$.

We acknowledge useful discussions with Conor Houghton and Nick
Manton. Our research is funded by EPSRC (PMS) and PPARC (RAB). The
parallel computations were perfomed at the National Cosmology
Supercomputing Centre in Cambridge.

\def\jnl#1#2#3#4#5#6{\hang{#1, {\it #4\/} {\bf #5}, #6 (#2).} }
\def\jnltwo#1#2#3#4#5#6#7#8{\hang{#1, {\it #4\/} {\bf #5}, #6; {\it
ibid} {\bf #7} #8 (#2).} } \def\prep#1#2#3#4{\hang{#1, #4.} }
\def\proc#1#2#3#4#5#6{{#1 [#2], in {\it #4\/}, #5, eds.\ (#6).} }
\def\prep#1#2#3#4{\hang{#1, #4 (#2).}}
\def\book#1#2#3#4{\hang{#1, {\it #3\/} (#4, #2).} }
\def\jnlerr#1#2#3#4#5#6#7#8{\hang{#1 [#2], {\it #4\/} {\bf #5}, #6.
{Erratum:} {\it #4\/} {\bf #7}, #8.} } \def\prl{Phys.\ Rev.\ Lett.}
\def\pr{Phys.\ Rev.}  \def\pl{Phys.\ Lett.}  \def\np{Nucl.\ Phys.}
\def\prp{Phys.\ Rep.}  \def\rmp{Rev.\ Mod.\ Phys.}  \def\cmp{Comm.\
Math.\ Phys.}  \def\mpl{Mod.\ Phys.\ Lett.}  \def\apj{Astrophys.\ J.}
\def\apjl{Ap.\ J.\ Lett.}  \def\aap{Astron.\ Ap.}  \def\cqg{Class.\
Quant.\ Grav.}  \def\grg{Gen.\ Rel.\ Grav.}  \def\mn{Mon.\ Not.\ Roy.\
Astro.\ Soc.}
\def\ptp{Prog.\ Theor.\ Phys.}  \def\jetp{Sov.\ Phys.\ JETP}
\def\jetpl{JETP Lett.}  \def\jmp{J.\ Math.\ Phys.}  \def\zpc{Z.\
Phys.\ C} \def\cupress{Cambridge University Press} \def\pup{Princeton
University Press} \def\wss{World Scientific, Singapore}
\def\oup{Oxford University Press}

\end{document}